\documentclass[sigconf]{acmart}
\usepackage[utf8]{inputenc}
\usepackage{graphicx}
\usepackage{pgfplots}
\usepackage{subcaption}
\pgfplotsset{compat=1.17}
\usepackage{multirow}
\usepackage{pgfplotstable}
\usetikzlibrary{pgfplots.statistics}

\usepackage{listings}
\definecolor{codegreen}{rgb}{0,0.6,0}
\definecolor{codegray}{rgb}{0.5,0.5,0.5}
\definecolor{codepurple}{rgb}{0.58,0,0.82}
\definecolor{backcolour}{rgb}{0.95,0.95,0.92}
\usepackage[most]{tcolorbox}
\lstdefinestyle{mystyle}{
    backgroundcolor=\color{backcolour},   
    commentstyle=\color{codegreen},
    keywordstyle=\color{magenta},
    numberstyle=\tiny\color{codegray},
    stringstyle=\color{codepurple},
    basicstyle=\ttfamily\scriptsize,
    breakatwhitespace=false,         
    breaklines=true,                 
    captionpos=b,                    
    keepspaces=true,
    numbers=left,                    
    numbersep=5pt,                  
    showspaces=false,                
    showstringspaces=false,
    showtabs=false,                  
    tabsize=2
}
\AtBeginDocument{%
  }

\setcopyright{acmlicensed}
\copyrightyear{2025}
\acmYear{2025}
\acmDOI{XXXXXXX.XXXXXXX}
\acmConference[EASE 2025]{The 29th International Conference on Evaluation and Assessment in Software Engineering}{17–20 June, 2025}{Istanbul, Turkey}
\acmISBN{978-1-4503-XXXX-X/2018/06}

\begin{document}

\title{Paradigm shift on Coding Productivity Using GenAI}

\author{Liang Yu}
\email{liang.yu@bth.se}
\orcid{0000-0001-5949-1375}
\authornotemark[1]
\affiliation{%
  \institution{Blekinge Institute of Technology}
  \city{Karlskrona}
  \state{Blekinge}
  \country{Sweden}
}

\renewcommand{\shortauthors}{Yu et al.}

\begin{abstract}
Generative AI (GenAI) applications are transforming software engineering by enabling automated code co-creation.
However, empirical evidence on GenAI's productivity effects in industrial settings remains limited.
This paper investigates the adoption of GenAI coding assistants (e.g., Codeium, Amazon Q) within telecommunications and FinTech domains.
Through surveys and interviews with industrial domain-experts, we identify primary productivity-influencing factors, including task complexity, coding skills, domain knowledge, and GenAI integration.
Our findings indicate that GenAI tools enhance productivity in routine coding tasks (e.g., refactoring and Javadoc generation) but face challenges in complex, domain-specific activities due to limited context-awareness of codebases and insufficient support for customized design rules.
We highlight new paradigms for coding transfer, emphasizing iterative prompt refinement, immersive development environment, and automated code evaluation as essential for effective GenAI usage.
\end{abstract}



\maketitle

\section{Introduction}
\label{sec:introduction}
\begin{figure}[ht]
    \centering
    \includegraphics[width=1\linewidth]{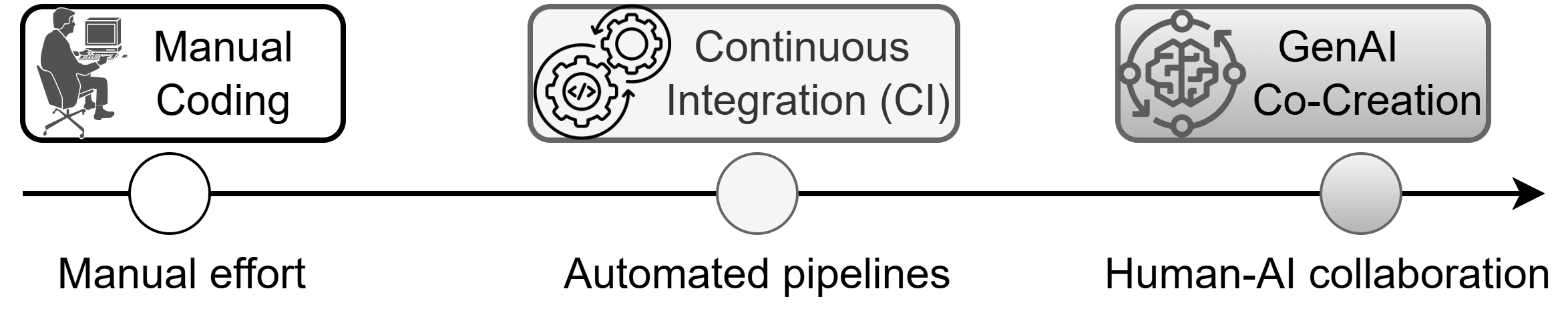}
    \caption{Progress in software development practices}
    \Description{Manual to automated and co-creation progress in software development practices}
    \label{fig:paradigm}
\end{figure}

Software development practices have progressed from manual coding toward increasingly automated and intelligent workflows~\cite{coutinho2024role}, as shown in Figure~\ref{fig:paradigm}.

A key milestone in this progress was the adoption of Continuous Integration (CI)~\cite{klotins2023continuous}.
CI automates verification and validation pipelines and offers fast feedback to reduce integration overhead and accelerate the coding cycle.
Building on CI pipelines for efficiency, \textit{Generative Artificial Intelligence} (GenAI)~\cite{gupta2024adoption} has emerged as a next step, leveraging Large Language Models (LLMs)~\cite{li2023large} to generate both text~\cite{vaithilingam2022expectation} and code~\cite{le2022coderl}.
Whereas CI improves code quality by rapidly detecting human errors, GenAI promises to expand automation into \textit{co-creation}, enabling developers to offload specific coding tasks to AI assistants (e.g., \href{https://codeium.com}{Codeium}, \href{https://aws.amazon.com/q}{Amazon Q}).

Despite these technological advances, there is limited empirical evidence~\cite{coutinho2024role} about how GenAI coding assistants affect broader aspects of \textit{coding productivity}, such as problem-solving process, learning experiences, and overall code quality~\cite{canedo2019factors}.
To address this gap, we investigate the impact of GenAI coding assistants in industrial contexts, analyzing how they reshape development practices and what lessons emerge for practitioners while adopting AI-supported code generation.
We assume that users experience productivity gains with GenAI compared to traditional coding work without GenAI, and domain-specific knowledge in using GenAI is related to the productivity impacts.

The \textit{main contributions} of this study are: I) Empirical data on GenAI coding tools’ impact on coding productivity; II) Practical lessons while using GenAI coding assistants.

The remainder of this paper is structured as follows.
Section~\ref{sec:background} introduces background information and related work.
Section~\ref{sec:methodology} outlines the research methodology.
Section~\ref{sec:results} reports study findings.
Finally, Section~\ref{sec:conclusions} concludes the study results and future research.

\section{Related work}
\label{sec:background}

While automated code generation has been explored for decades, recent approaches utilizing LLM-based models have opened new opportunities.
Omidvar et al.~\cite{omidvar2024evaluating} studied LLM-based code migration, highlighting the importance of a human-AI partnership in complex refactoring tasks.
Similarly, Heng et al.~\cite{heng2024comparing} showed that LLM-driven code generation could support learning new frameworks (e.g., Flutter), indicating potential productivity gains and educational benefits.

Other studies focus on the empirical impact of AI-based code assistants in professional environments.
Gon{\c{c}}alves et al.~\cite{goncalves2024assessment} conducted an empirical assessment of GitHub Copilot, revealing gains in coding speed and noting concerns regarding code correctness.
Corso et al.~\cite{corso2024assessing} extended this line of work by measuring the adequacy of AI-based code assistants in method-generation tasks, underscoring the crucial need to manage AI-generated technical debt.
Despite these contributions, the literature lacks evidence concerning qualitative dimensions of \textit{productivity}, such as industrial engineers' quality perceptions and design rules sharing across teams.
Our work addresses this gap by investigating industrial settings, offering insights into how GenAI tools reshape modern software development.

\subsection{Definitions}\label{sec:definitions}
\textbf{Productivity:} Inspired by Gon{\c{c}}alves et al.~\cite{goncalves2024assessment}, we define productivity as a combination of \textit{user satisfaction, coding tasks, acquired skills/knowledge from using tools, and quality perceptions of tools' outputs} rather than merely lines of code completed or numbers of bugs reduced.
Here, coding tasks refer to activities such as implementing features, refactoring code, debugging, writing tests, code reviews, and documentation.

\section{Research Methodology}
\label{sec:methodology}
We conducted a multi-case study following Runeson et al.'s guidelines~\cite{runeson2009guidelines}.
Survey and semi-structured interview methods were used to extract in-depth insights into the uses of GenAI coding assistants.

\begin{figure}[ht]
    \centering
    \includegraphics[width=\linewidth]{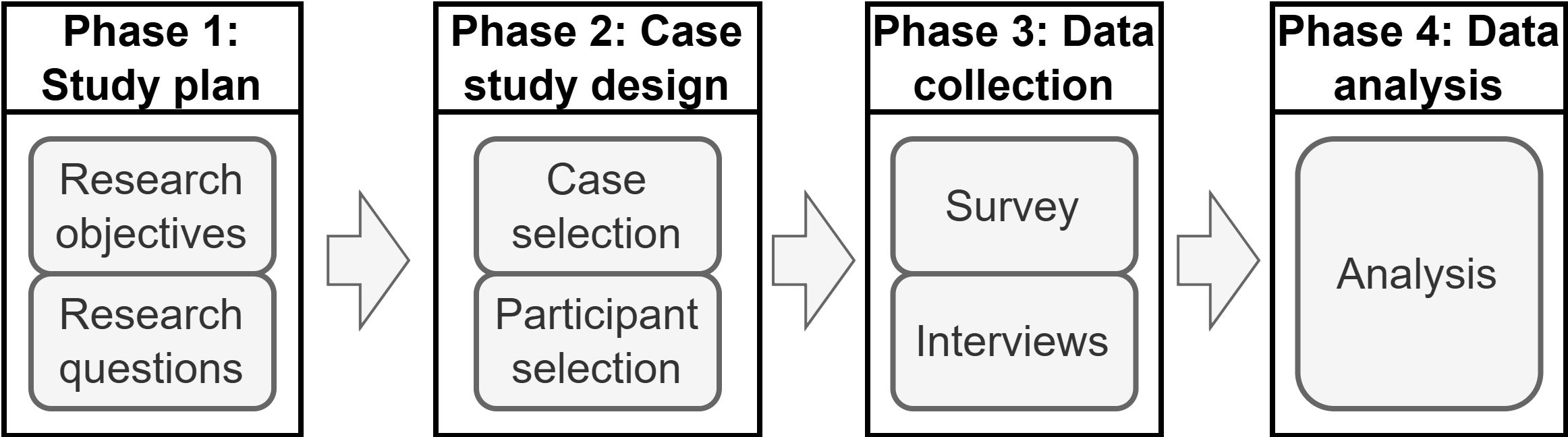}
    \caption{Research process overview}
    \label{fig:research_process}
    \Description{The Research process is in four phases}
\end{figure}

An overview of the research process is shown in Figure~\ref{fig:research_process}.
The process includes four phases: Study plan, Case study design, Data collection, and Data analysis.
We present details of these phrases from Section~\ref{subsec:study_plan} to Section~\ref{subsec:dataanalysis}.

\subsection{Phase 1: Study plan}\label{subsec:study_plan}
In this phase, we formulated research objectives and questions.

\subsubsection{Research objectives}
We aim to investigate insights about coding productivity by exploring the practices of using GenAI coding assistant tools in industrial settings.

\subsubsection{Research questions}
\label{subsec:RQs}
We formulated two research questions:
\begin{itemize}
    \item \textbf{RQ1:} What factors impact practitioners’ productivity when using GenAI coding assistants?
    \item \textbf{RQ2:} What lessons emerge from practitioners while adopting GenAI coding assistants?
\end{itemize}


\subsection{Phase 2: Case study design}
This section presents the case and participant selections.

\subsubsection{Case selection}
\href{www.ericsson.com}{Ericsson} is our case company in this study.
It leads the way in using GenAI coding assistants in the telecommunications domain for software development teams.

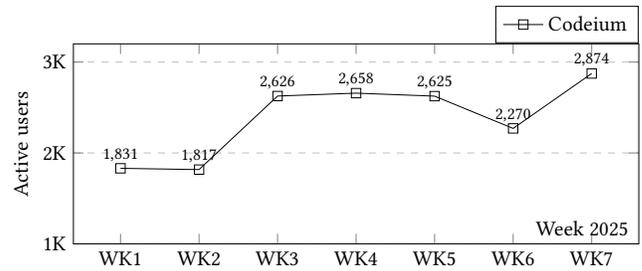
\begin{figure}[ht]
    \centering
    \resizebox{1\linewidth}{!}{
    \begin{tikzpicture}
    \begin{axis}[
        width=1\linewidth,
        height=3cm,
        scale only axis,
        xlabel={Week 2025},
        xlabel style={at={(0.9,0.15)}, anchor=north},
        ylabel={Active users},
        legend style={at={(1,1.19)}, anchor=north east, legend columns=-1},
        symbolic x coords={WK1,WK2,WK3,WK4,WK5,WK6,WK7},
        xtick=data,
        ymin=1000,
        ymax=3200,
        ytick={1000,2000,3000},
        yticklabels={1K,2K,3K},
        ymajorgrids=true,
        grid style=dashed
    ]
    
    \addplot[
        black,
        mark=square,
        nodes near coords,
        every node near coord/.append style={font=\footnotesize}
    ] 
    table[row sep=crcr] {
        dummy 1000\\ 
        WK1   1831\\
        WK2   1817\\
        WK3   2626\\
        WK4   2658\\
        WK5   2625\\
        WK6   2270\\
        WK7   2874\\
    };
    \addlegendentry{Codeium}
    \end{axis}
    \end{tikzpicture}%
    }
    \caption{Active users per week for Codeium}
    \label{fig:active_users}
    \Description{Active users per week for utilizing Codeium.}
\end{figure}

Coding assistants in the case company include Codeium, Amazon Q, Tabnine, and GitHub Copilot.
Recent statistics gathered by the first author, who works as an embedded developer within the company, indicate that Codeium has emerged as the most widely adopted tool.
As illustrated in Figure~\ref{fig:active_users}, the number of Codeium users grew by 57\% at the start of 2025.
Amazon Q ranks second, with approximately 200 active users, while Tabnine and GitHub Copilot currently have minimal or no usage.
A likely explanation for the low adoption of the latter tools is security concerns, such as a lack of support for on-premise deployment or risks that may unintentionally expose proprietary code to third parties.

\begin{table}[ht]
    \caption{Context information~\cite{petersen2009context} of the selected projects}
    \label{tab:project-contexts}
    \resizebox{0.9\linewidth}{!}{
    \begin{tabular}{|l|l|l|l|l|}
    \hline
    \multicolumn{2}{|l|}{\textbf{\begin{tabular}[c]{@{}l@{}}Context \end{tabular}}} &
      \textbf{Project A} &
      \textbf{Project B}\\ \hline
    \multirow{3}{*}{\textit{Project}} &
      \begin{tabular}[c]{@{}l@{}}Business\\ domain\end{tabular} &
      FinTech &
      Telecom \\ \cline{2-4} 
     &
      Product type &
      Finance services &
      \begin{tabular}[c]{@{}l@{}}Network \\provisioning\end{tabular} \\ \cline{2-4} 
     &
      \begin{tabular}[c]{@{}l@{}}Maturity\end{tabular} &
      Mature product &
      Mature product \\ \hline
    \multirow{1}{*}{\textit{Organization}} &
      \begin{tabular}[c]{@{}l@{}}number of\\ engineers\end{tabular} &
      \textgreater 500 &
      \textgreater 200 \\ \hline
    \textit{Development} &
      \begin{tabular}[c]{@{}l@{}}GenAI usage\end{tabular} &
      Yes &
      Yes \\ \hline
    \end{tabular}%
    }
\end{table}

We selected two projects, denoted as Project A and B, using purposive sampling~\cite{patton2002qualitative} from our existing industrial network.
At the time of the study, these two projects used GenAI coding assistants.
Table~\ref{tab:project-contexts} presents the contexts of the selected projects.
We started our research on the selected project in this study.
We plan to replicate the research on the other projects shortly.

Project A offers financial services (e.g., transactions, loans, and payments) that have successfully served customers for approximately two decades.
Project B is in the telecommunications domain, providing support services for network management.
Both projects are rapidly growing, and the need to use GenAI to scale their business grows.
Our study focused on the product development department in Sweden, which employs approximately 200 engineers.

\subsubsection{Participant selection}
Participants were selected using convenience sampling~\cite{runeson2009guidelines}, determined by the individuals' availability and willingness to participate in the study.

\begin{table}[H]
\caption{Participants from the selected Project A and B}
\label{tab:participants}
\resizebox{0.95\linewidth}{!}{%
\begin{tabular}{|l|c|c|c|}
\hline
\textbf{Job role} & \textbf{\begin{tabular}[c]{@{}c@{}}NO. of \\ participants\end{tabular}} & \textbf{\begin{tabular}[c]{@{}c@{}}Work experience \\ (years)\end{tabular}} & \textbf{Project} \\ \hline
 Software developer        & 9                                                                       & 3 - 20                                                                         & A (5), B (4)   \\ \hline
Software architect        & 5                                                                       & 7 - 26                                                                         & A (3), B (2)   \\ \hline
DevOps engineer        & 4                                                                       & 5 - 14                                                                         & A (2), B (2)   \\ \hline
\end{tabular}%
}
\end{table}

As shown in Table~\ref{tab:participants}, 18 participants were chosen.
All participants have experience in using either Codeium or Amazon Q.
Participant roles included software developers, software architects, and DevOps engineers.
Their working experience ranges from three to 26 years.

\subsection{Data collection}
\label{subsec:design}
We developed a survey with a questionnaire to gather feedback on using GenAI coding assistant tools.
The survey included: I) demographics, roles, and years of experience; II) tool usage (e.g., frequency and duration of use); III) Likert-scale items, e.g., perceptions of satisfaction and tool integration.

Following the survey, we conducted 17 interviews with participants to gain further practical insights.
The interviews were conducted in a mixed format, with onsite and online meetings, depending on participants' availability.
Each interview lasted between 33 and 45 minutes.
All authors participated in designing the survey and interviews through four rounds of meetings and refinements.

\subsubsection{Data Availability}
The survey results and extracted data are available at \href{https://figshare.com/s/1d36c6cec3d72c2c3e89}{Figshare}.
\acmDataLink{https://figshare.com/s/1d36c6cec3d72c2c3e89}

\subsubsection{Measuring productivity}\label{sec:productivity_measures}
In line with the definition presented in Section~\ref{sec:definitions}, we asked participants to rate their satisfaction with AI-generated suggestions on a five-point Likert scale (1=very low, 5=very high).
We gathered estimates of time saved by comparing coding effort with and without GenAI for everyday tasks (e.g., writing tests and code refactoring).
We examined interview transcripts for examples of newly acquired techniques or clarifications of existing knowledge, representing learning gains.
Finally, to assess the quality of GenAI tools' outputs, we asked how often participants needed to revise AI-generated code, using a five-point scale and follow-up interview probes for detailed explanations.

\subsection{Data analysis}
\label{subsec:dataanalysis}
We first computed frequencies and means for the Likert-scale data (e.g., satisfaction, integration effort levels), enabling us to gauge overall trends. 
Next, we carried out a step-by-step thematic analysis~\cite{cruzes2011recommended} on the collected qualitative data.

\begin{figure}[htbp]
    \centering
    \includegraphics[width=0.9\linewidth]{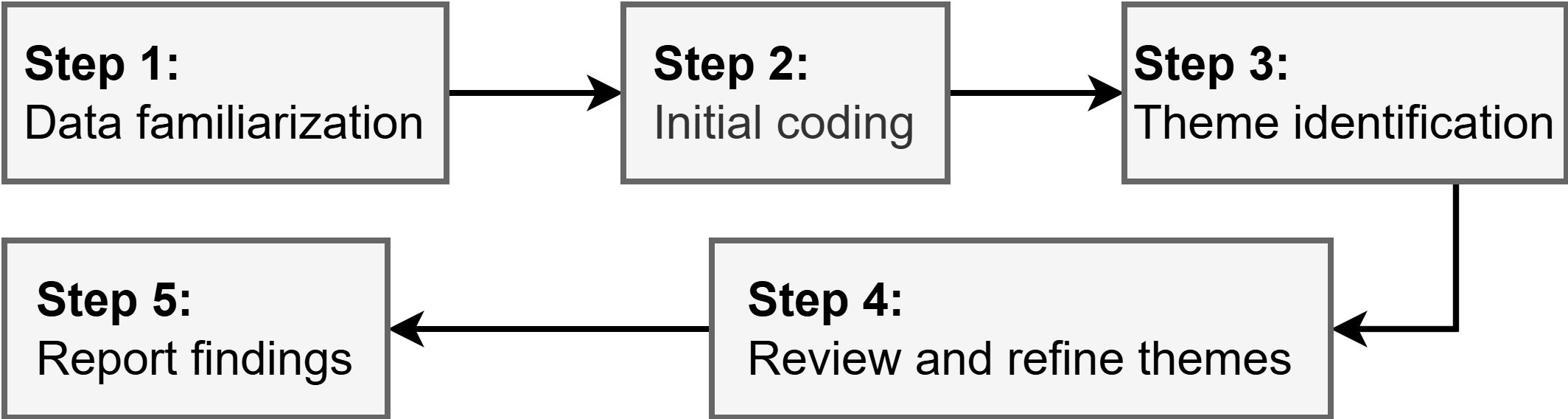}
    \caption{Data analysis steps}
    \label{fig:data_analysis}
    \Description{Data analysis based on collected data from surveys and interviews.}
\end{figure}

Figure~\ref{fig:data_analysis} illustrates the data analysis steps of our study.
The first author performed the data analysis, and all authors reviewed and discussed the results.

\textbf{Step 1: Data familiarization.}
We read both survey open-ended responses and interview transcripts sentence by sentence to note initial impressions.

\textbf{Step 2: Initial coding.}
We inductively labeled codes in the sentences or phrases created in Step 1.
Codes refer to words or terms that could represent potential main subjects, for example, ``user satisfaction'', ``productivity'', and ``domain knowledge.''

\textbf{Step 3: Theme identification.}
We grouped similar codes into higher-level themes.
For example,  ``types of tasks,'' ``human oversight,'' ``IDE integration,'' and ``Codebase context''.

\textbf{Step 4: Review and refine themes.}
Firstly, we reviewed the relationship between themes, i.e., whether higher-level themes were grouping lower-level themes appropriately.
Secondly, we updated duplicate themes and required changes in the names of the themes.
Thirdly, we merged and refined themes where necessary, ensuring their consistency.

\textbf{Step 5: Report findings.}
We utilized our refined themes to identify recurring usage patterns of GenAI coding assistants, which informed our main findings.

\section{Results}
\label{sec:results}
This section presents our findings to answer the research questions.

\subsection{Results of RQ1: What factors impact practitioners' productivity when using AI coding assistants?}

According to study participants, the duration of using GenAI coding assistants ranged from one month to over six months.
Based on the collected data, we have synthesized five factors that affect practitioners' productivity, as shown in Table~\ref{tab:factors}.
Our findings for these factors are presented from Section~\ref{sec:tasks-types} to Section~\ref{sec:tool-familiar}.

\begin{table}[htp]
\caption{Factors that impact practitioners' productivity}
\label{tab:factors}
\resizebox{0.9\columnwidth}{!}{%
\begin{tabular}{|l|l|}
\hline
\textbf{Factor name} & \textbf{Description}                                                                                                                                    \\ \hline
Task types           & \begin{tabular}[c]{@{}l@{}}Task types refer to code generation,\\ refactoring, reviews, explanation \\and test generation.\end{tabular} \\ \hline
Coding skills    & \begin{tabular}[c]{@{}l@{}}Skills of writing and debugging code.\end{tabular}                                                                     \\ \hline
Domain knowledge     & \begin{tabular}[c]{@{}l@{}}Expertise in a specific field or\\ programming experience.\end{tabular}                                                          \\ \hline
IDE integration     & \begin{tabular}[c]{@{}l@{}}Connecting a tool with an integrated\\ development environment (IDE).\end{tabular}                                                       \\ \hline
Tool familiarity     & \begin{tabular}[c]{@{}l@{}}Learning/training on how to \\ use a specific software.\end{tabular}                                                       \\ \hline
\end{tabular}%
}
\end{table}

\subsubsection{Task types}\label{sec:tasks-types}
Different coding tasks result in different productivity perceptions.
These tasks include code generation, refactoring, review, and test creation.
60\% of participants cited reduced effort for code refactoring (e.g., functions and variables updates) and review (e.g., code explanation) tasks.
For example, several developers noted: \textit{``It generates repetitive blocks of code and similar variable definitions easily and quickly. It avoids the pain of typing again.''}

However, the capacity of GenAI coding tools for complex coding tasks (e.g., code/test generation for Gradle modules in JAVA projects) was limited.
Participants stated: \textit{``If you ask the tool to make changes to the whole file, it will often start to rewrite code unrelated to the task you asked it to solve.''}

\begin{figure}[htp]
    \centering
    \includegraphics[width=\columnwidth]{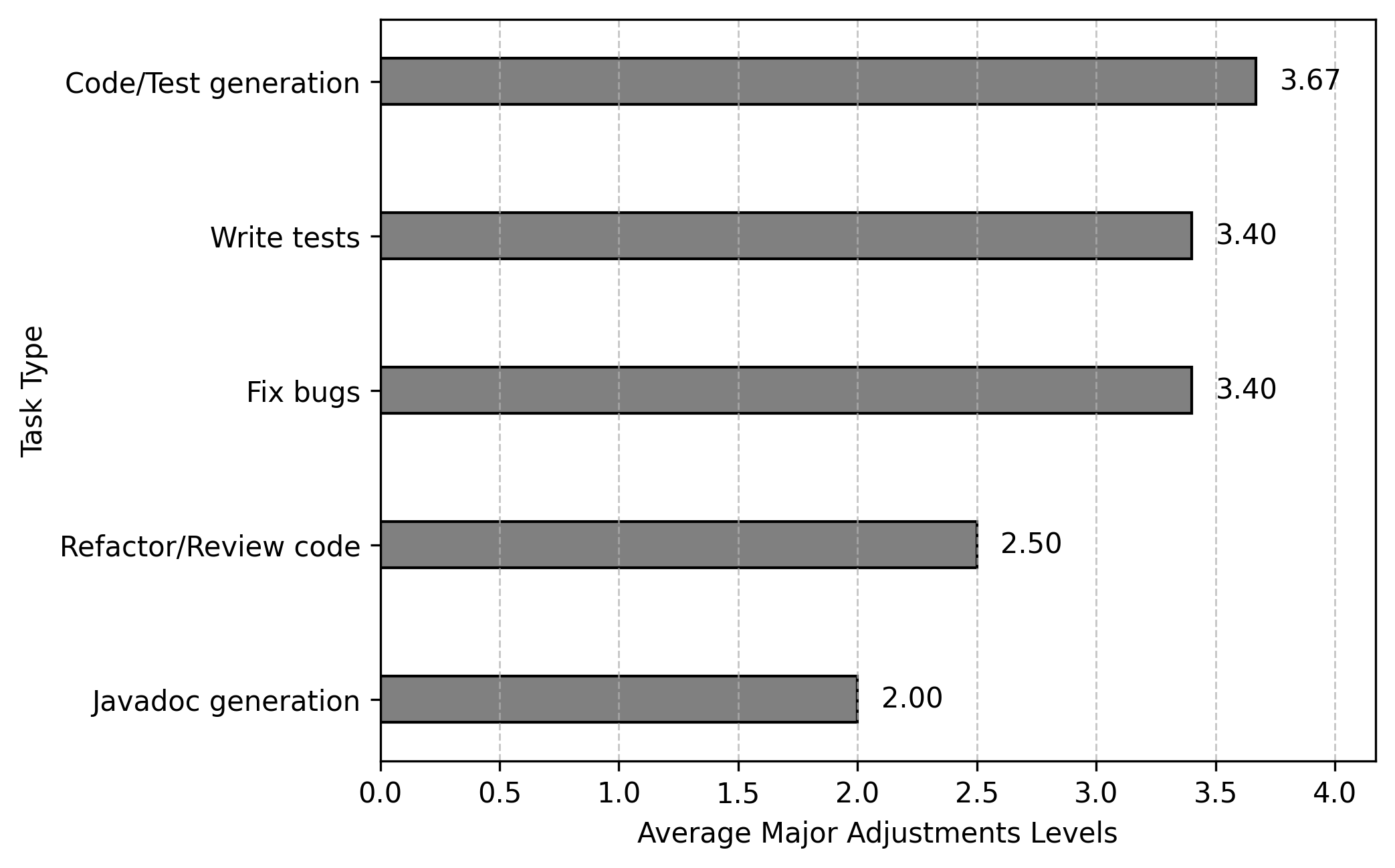}
    \caption{Average major adjustments required for different types of tasks based on AI-generated outputs}
    \Description{Average major adjustments required for different types of tasks}
    \label{fig:major-adjustments}
\end{figure}

Regarding the quality of AI-generated code, we evaluated mean primary adjustment levels across the identified task types.
As shown in Figure~\ref{fig:major-adjustments}, ``Code/Test generation'' has the highest level at about 3.67, suggesting that creating or generating code involves the most significant revisions.
As participants criticized: \textit{``Code generation of these tools is not good enough compared to OpenAI models,''} which indicates concerns that not all AI models are equally proficient.
Both ``Write tests'' and ``Fix bugs'' show similar levels at around 3.4.
``Refactor/Review code'' has a moderate level of 2.5, indicating fewer major changes.
For example, several participants reviewed the quick AI-generated code suggestions: \textit{``You always need to examine what it suggests.''}
``Javadoc generation'' stands lowest at 2, indicating minimal required adjustments.

\subsubsection{Coding skills}
Participants commented that with more coding experience gained, less productivity was perceived while using AI coding assistants.
One possible reason could be that with more coding experience, users know exactly how to complete a task without retrieving or acquiring external information.
Since processing information from AI coding assistant tools consumes time.
Like a participant complained: \textit{``Lack of insight into what code the plugin has in its context, what I should expect, and when I should not waste my time trying to get the chat to understand.''}

However, users with rich coding experience may neglect recent/advanced knowledge they may not be familiar with.
As participants complemented: \textit{``The solution suggested by AI tools might not be one that you would have considered before, and then you can learn new ways of doing things.''}

\subsubsection{Domain knowledge} Domain-experts with a better understanding of contextualized codebase information are more capable of judging whether AI-generated or human-developed solutions.

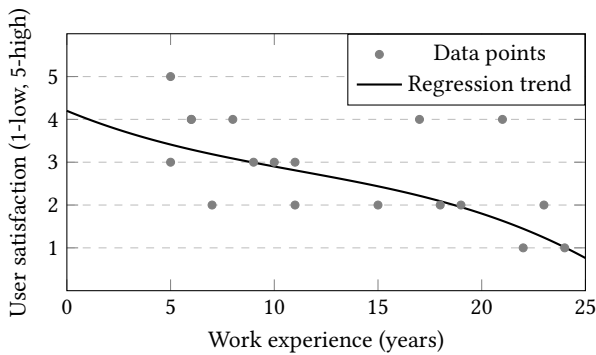
\begin{figure}[ht]
    \centering
    \begin{tikzpicture}
        \begin{axis}[
            width=\linewidth,
            height=5cm,
            xlabel={Work experience (years)},
            ylabel={User satisfaction (1-low, 5-high)},
            xmin=0, xmax=25,
            ymin=0, ymax=6,
            xtick={0,5,10,15,20,25},
            ytick={1,2,3,4,5},
            ymajorgrids=true,
            grid style=dashed,
            legend style={at={(1,1)}, anchor=north east},
        ]

        \addplot[
            only marks,
            mark=*,
            mark size=1.5pt,
            gray
        ] 
        coordinates {
            (5,5) (5,3) (6,4) (23,2) (24,1) (15,2) (18,2) (17,4) 
            (6,4) (21,4) (9,3) (10,3) (11,3) (22,1) (8,4) (7,2) (11,2) (19,2)
        };
        \addlegendentry{Data points}

        \addplot[
            black,
            thick,
            smooth,
            domain=0:25,
            samples=100
        ]
        {4.2 - 0.2*x + 0.01*x^2 - 0.0003*x^3};
        \addlegendentry{Regression trend}

        \end{axis}
    \end{tikzpicture}
    \caption{The relation between user satisfaction and working experience with regression trend}
    \label{fig:regression-curve}
    \Description{The relation between user satisfaction and working experience with regression trend.}
\end{figure}

As shown in Figure~\ref{fig:regression-curve}, the regression trend indicates that the higher the user's ability to validate AI-generated code, the lower the perceived satisfaction.
On the contrary, for users with limited knowledge of how to validate AI-generated code, AI tools provide higher potential value but also introduce risks without proper validation.

Additionally, common issues mentioned included over-aggressive suggestions and mismatches between the suggested code and the user's actual expectations.
Participants highlighted their frustration: \textit{``Wrong auto-completion code suggestions''} and \textit{``The tool does not have a full understanding of my code repository.''}

\subsubsection{IDE integration}
As shared by participants, a simple user interface (UI) or quick enabling/disabling of code suggestions was seen as a productivity booster.

\begin{figure}[ht]
    \centering
    \begin{tikzpicture}
        \begin{axis}[
            width=\linewidth,
            height=5cm,
            xlabel={IDE integration effort levels},
            ylabel={Number of participants},
            symbolic x coords={1 (very easy), 2, 4, 5 (very difficult)},
            xtick=data,
            ybar,
            ymin=0,
            ymajorgrids=true,
            grid style=dashed,
            nodes near coords,
            bar width=10pt,
        ]

        \addplot[
            fill=gray!50
        ] coordinates {
            (1 (very easy), 11)
            (2, 5)
            (4, 2)
            (5 (very difficult), 0)
        };

        \end{axis}
    \end{tikzpicture}
    \caption{The effort levels while integrating GenAI coding tools into integrated development environments (IDEs)}
    \label{fig:integration-effort}
    \Description{The effort levels while integrating GenAI coding assistant tools into integrated development environments.}
\end{figure}
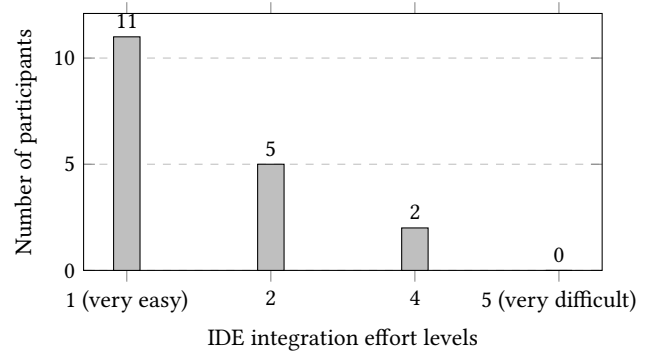

Figure~\ref{fig:integration-effort} presents the distribution of effort levels reported by participants while integrating GenAI coding assistant tools into their IDEs.
The \textit{x}-axis represents different levels of integration effort, ranging from Level 1 (very easy) to Level 5 (very difficult), while the \textit{y}-axis shows the number of participants at each level.

Study results show that most participants (11 out of 18) found the integration process very easy (Level 1), suggesting that GenAI coding assistant tools are straightforward to integrate into existing development environments.
Additionally, five participants reported an effort as Level 2, while two participants rated the effort as Level 4.
Notably, no participants reported the highest difficulty (Level 5 – very difficult), further reinforcing that the integration process is not highly complex.

These findings suggest that GenAI coding assistants require minimal setup effort for most users, potentially reducing adoption barriers and enabling smoother workflows within IDEs.

\subsubsection{Tool familiarity}\label{sec:tool-familiar}
Study results indicate that as users utilized the AI coding assistant tools more frequently, they gained better results and learned to \textit{bounce ideas} effectively.

\begin{figure}[ht]
    \centering
    \includegraphics[width=\columnwidth]{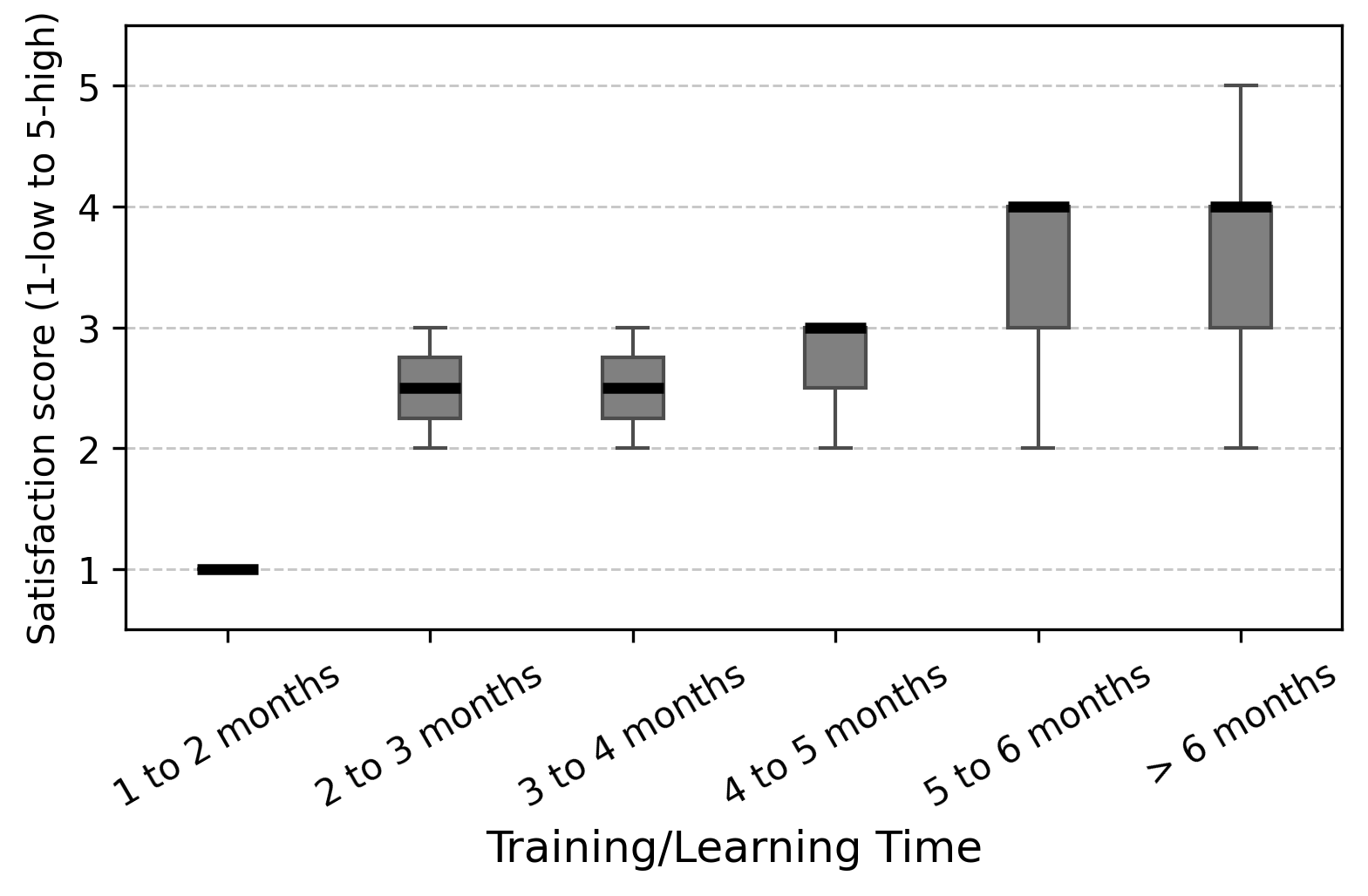}
    \caption{Median satisfaction scores based on the usage/training time of coding assistants}
    \label{fig:training-satisfaction}
    \Description{Relationship between tool usage time and satisfaction score.}
\end{figure}

Figure~\ref{fig:training-satisfaction} shows median satisfaction scores for using or learning AI coding assistants across different periods.
The figure reveals that users with limited experience (1 to 2 months) report the lowest satisfaction, whereas satisfaction generally increases with extended usage.
Users with five to six months of experience exhibit a higher median satisfaction score of around 4.

For those exceeding six months, satisfaction remains high but exhibits variability, suggesting that prolonged use leads to mixed experiences.
This finding indicates that familiarity with the tool improves satisfaction within the first six months.
However, new challenges or limitations emerge as users become more accustomed to specific coding assistant tools, potentially reducing satisfaction.

\subsubsection{Summary}
In short, our findings show that domain knowledge and the GenAI tool's training interact with how productivity is realized.
Partial productivity gains, such as time savings of about 13\%, have been confirmed by study participants.
However, these gains are situational, primarily seen in boilerplate generation rather than complex implementation tasks, such as creating a new Gradle module in a codebase.

\subsection{Results of RQ2: What lessons emerge from practitioners while adopting GenAI coding assistants?}

\begin{tcolorbox}[colback=gray!5!white,colframe=gray!75!black,title=Lesson learned 1]
  The type of coding tasks matters when measuring users' productivity.
\end{tcolorbox}

Study participants reported that GenAI coding assistants excel at generating boilerplate code.
For instance, a DevOps developer commented: 
\textit{``It auto-suggests repeating blocks of code right away, which saves me a lot of tedious typing.''}

However, more complex tasks, particularly those requiring specialized domain knowledge or deeper architectural considerations, rarely showed comparable benefits.
One architect with fifteen years of experience in Project B explained:
\textit{``If it is a completely new functionality that needs deep understanding of our Gradle plugins, it takes longer to guide the AI than to just do it myself.''}
Two possible reasons were mentioned for this shortfall:
I) The GenAI’s suggestions often lacked context about custom frameworks (e.g., Junit5-based function test framework) or legacy code structures (e.g., monolith components); II) repeated re-prompting was needed to correct inaccuracies, undermining any time savings.

In addition, a developer found that, when exploring an unfamiliar payments API,
\textit{``GenAI assistants (e.g., Codeium) actually had newer OpenAPI examples, so I only had to make a few small tweaks.''}
This successful case underscores how GenAI’s broader training data can sometimes yield direct gains for complex tasks, provided the API in question is well-represented in its knowledge base.

\begin{tcolorbox}[colback=gray!5!white,colframe=gray!75!black,title=Lesson learned 2]
  AI coding assistants allow users to focus more on development tasks without leaving their IDEs.
\end{tcolorbox}
Participants praised how Codeium and Amazon Q integrate seamlessly into their development environments, minimizing the need to switch between web searches.
Several developers repeated:
\textit{``I no longer have to open a browser and check StackOverflow.
The assistant suggests relevant snippets in my IDE.''}
Participants stated that this single environment helps them maintain concentration, particularly when working with multiple files in large repositories.
Although some participants mentioned limitations with the GenAI’s contextual accuracy, most agreed that staying in the IDE reduces distraction and promotes an immersive development environment.

\begin{tcolorbox}[colback=gray!5!white,colframe=gray!75!black,title=Lesson learned 3]
  Context-related suggestions are challenging while using AI coding assistants.
\end{tcolorbox}
While AI coding assistants offer support for routine tasks, they often struggle with providing context-aware suggestions, which is a confirmation of Nam et al.'s \cite{nam2024using} findings.
As participants noted: \textit{``The tool sometimes failed to grasp the nuances of my project's specific framework, leading to suggestions that required significant manual adjustment.''}
Similarly, a participant complained that \textit{``I often found that the generated code was too generic, missing the unique context of our specialized codebase.''}

Even when partial code context is provided through IDE extensions, the models sometimes overlook important files, dependencies, or architectural conventions beyond their immediate context window.
This leads to generic recommendations that developers must extensively rework, which reduces the initial efficiency gains of using GenAI coding assistants.

\begin{tcolorbox}[colback=gray!5!white,colframe=gray!75!black,title=Lesson learned 4]
  Continuous and iterative refinement of prompts may result in more accurate suggestions.
\end{tcolorbox}
Participants discovered that prompt phrasing and incremental clarifications influenced the accuracy of AI-generated code.
A software architect remarked:
\textit{``The first suggestion was often incomplete, but after I clarified the requirement step by step, the tool finally got it right.''}
Similarly, a developer shared:
\textit{``I used to give a long prompt once, and the output was so-so. Now, I break the task into smaller pieces and guide the AI iteratively. That works much better.''}

By narrowing the scope or specifying constraints (e.g., target frameworks, data types), practitioners could make the GenAI tools produce code closer to their expectations.
However, frequent re-prompting can become time-consuming if the user must repeatedly correct the GenAI assistant’s misunderstandings, particularly in complex domains.
This shows the importance of balancing iterative guidance against potential overhead.

\begin{tcolorbox}[colback=gray!5!white,colframe=gray!75!black,title=Lesson learned 5]
  Customized design rules are not supported by GenAI coding assistants.
\end{tcolorbox}
Study participants complained that neither Codeium nor Amazon Q offers robust support for customized design rules that can be configured in IDEs.
As a result, the AI-generated code reflected general coding patterns rather than the project’s unique standards.

Design guidelines are often organization-specific and involve layered constraints (e.g., architecture, security, performance).
Current GenAI models are not tuned to interpret and apply such interdependent rules out of the box.
Consequently, practitioners had to perform manual adjustments to align their custom rule sets, thus reducing the overall benefit of AI-generated code.

\begin{tcolorbox}[colback=gray!5!white,colframe=gray!75!black,title=Lesson learned 6]
  Quality evaluation of AI-generated code is required.
\end{tcolorbox}
While study participants saw the convenience of AI-generated suggestions, they expressed concerns about verifying code performance, security, and maintainability at scale.
As an architect commented:
\textit{``We can do code reviews, but as soon as the AI starts generating large chunks of logic, manual oversight isn't enough.
We need automated checks.''}
Given this, scalable and automated quality evaluation frameworks are needed for AI-generated code.
Future work in this area is expected.

\subsection{Limitations}
Potential limitations include the small sample size, which may not generalize across diverse software domains.
Further, some participants may have used multiple GenAI tools (e.g., ChatGPT) outside our scope.
To mitigate bias, we anonymized responses and encouraged honest feedback.

\balance
\section{Conclusions}
\label{sec:conclusions}
This study investigated how practitioners in two industrial projects adopted GenAI coding assistants (Codeium and Amazon Q) and examined the resulting impact on their productivity.
Study results revealed that while GenAI tools accelerate everyday tasks such as code refactoring and Javadoc generation, they offer less consistent benefits for complex or domain-intensive work.

We learned that repeated re-prompting, limited domain-specific knowledge, and the lack of integrated design rules create barriers to time savings.
Conversely, we found that iterative refinement of prompts can improve output quality, but this process requires additional effort and oversight to be effective.
Our findings highlight that the productivity impact of GenAI is not uniform.
Domain knowledge, skills, and tooling integration influence the values of GenAI coding assistants in practice.

Future work includes replicating our multi-case approach in additional domains and expanding the sample size to complement and enrich our findings.
Further research on contextualized codebases and customized design rules would increase the usage of GenAI.

\begin{acks}
We acknowledge support from the Knowledge Foundation through the S.E.R.T. Research Profile Project (research profile grant 20180010).
Thanks also to Jon Påhls, Martin Blom, and Ulf Santesson, who supported us throughout this study.
\end{acks}

\bibliographystyle{ACM-Reference-Format}
\bibliography{manuscripts}


\begin{thebibliography}{16}


\ifx \showCODEN    \undefined \def \showCODEN     #1{\unskip}     \fi
\ifx \showISBNx    \undefined \def \showISBNx     #1{\unskip}     \fi
\ifx \showISBNxiii \undefined \def \showISBNxiii  #1{\unskip}     \fi
\ifx \showISSN     \undefined \def \showISSN      #1{\unskip}     \fi
\ifx \showLCCN     \undefined \def \showLCCN      #1{\unskip}     \fi
\ifx \shownote     \undefined \def \shownote      #1{#1}          \fi
\ifx \showarticletitle \undefined \def \showarticletitle #1{#1}   \fi
\ifx \showURL      \undefined \def \showURL       {\relax}        \fi
\providecommand\bibfield[2]{#2}
\providecommand\bibinfo[2]{#2}
\providecommand\natexlab[1]{#1}
\providecommand\showeprint[2][]{arXiv:#2}

\bibitem[Canedo and Santos(2019)]%
        {canedo2019factors}
\bibfield{author}{\bibinfo{person}{Edna~Dias Canedo} {and} \bibinfo{person}{Giovanni~Almeida Santos}.} \bibinfo{year}{2019}\natexlab{}.
\newblock \showarticletitle{Factors affecting software development productivity: An empirical study}. In \bibinfo{booktitle}{\emph{Proceedings of the XXXIII Brazilian Symposium on Software Engineering}}. \bibinfo{pages}{307--316}.
\newblock


\bibitem[Corso et~al\mbox{.}(2024)]%
        {corso2024assessing}
\bibfield{author}{\bibinfo{person}{Vincenzo Corso}, \bibinfo{person}{Leonardo Mariani}, \bibinfo{person}{Daniela Micucci}, {and} \bibinfo{person}{Oliviero Riganelli}.} \bibinfo{year}{2024}\natexlab{}.
\newblock \showarticletitle{Assessing {AI}-{Based} {Code} {Assistants} in {Method} {Generation} {Tasks}}. In \bibinfo{booktitle}{\emph{Proceedings of the 2024 {IEEE}/{ACM} 46th {International} {Conference} on {Software} {Engineering}: {Companion} {Proceedings}}}. \bibinfo{publisher}{ACM}, \bibinfo{address}{Lisbon Portugal}, \bibinfo{pages}{380--381}.
\newblock
\showISBNx{9798400705021}
\href{https://doi.org/10.1145/3639478.3643122}{doi:\nolinkurl{10.1145/3639478.3643122}}


\bibitem[Coutinho et~al\mbox{.}(2024)]%
        {coutinho2024role}
\bibfield{author}{\bibinfo{person}{Mariana Coutinho}, \bibinfo{person}{Lorena Marques}, \bibinfo{person}{Anderson Santos}, \bibinfo{person}{Marcio Dahia}, \bibinfo{person}{Cesar Fran{\c{c}}a}, {and} \bibinfo{person}{Ronnie de Souza~Santos}.} \bibinfo{year}{2024}\natexlab{}.
\newblock \showarticletitle{The role of generative ai in software development productivity: A pilot case study}. In \bibinfo{booktitle}{\emph{Proceedings of the 1st ACM International Conference on AI-Powered Software}}. \bibinfo{pages}{131--138}.
\newblock


\bibitem[Cruzes and Dyba(2011)]%
        {cruzes2011recommended}
\bibfield{author}{\bibinfo{person}{Daniela~S Cruzes} {and} \bibinfo{person}{Tore Dyba}.} \bibinfo{year}{2011}\natexlab{}.
\newblock \showarticletitle{Recommended steps for thematic synthesis in software engineering}. In \bibinfo{booktitle}{\emph{2011 International Symposium on Empirical Software Engineering and Measurement}}. IEEE, \bibinfo{pages}{275--284}.
\newblock


\bibitem[Gonçalves and Gonçalves(2025)]%
        {goncalves2024assessment}
\bibfield{author}{\bibinfo{person}{Carlos~Adriano Gonçalves} {and} \bibinfo{person}{Célia~Talma Gonçalves}.} \bibinfo{year}{2025}\natexlab{}.
\newblock \showarticletitle{Assessment on the Effectiveness of GitHub Copilot as a Code Assistance Tool: An Empirical Study}.
\newblock \bibinfo{journal}{\emph{Lecture Notes in Computer Science (including subseries Lecture Notes in Artificial Intelligence and Lecture Notes in Bioinformatics)}}  \bibinfo{volume}{14969 LNAI} (\bibinfo{year}{2025}), \bibinfo{pages}{27 – 38}.
\newblock
\href{https://doi.org/10.1007/978-3-031-73503-5_3}{doi:\nolinkurl{10.1007/978-3-031-73503-5_3}}
\newblock
\shownote{Cited by: 0}.


\bibitem[Gupta et~al\mbox{.}(2024)]%
        {gupta2024adoption}
\bibfield{author}{\bibinfo{person}{Ruchi Gupta}, \bibinfo{person}{Kiran Nair}, \bibinfo{person}{Mahima Mishra}, \bibinfo{person}{Blend Ibrahim}, {and} \bibinfo{person}{Seema Bhardwaj}.} \bibinfo{year}{2024}\natexlab{}.
\newblock \showarticletitle{Adoption and impacts of generative artificial intelligence: Theoretical underpinnings and research agenda}.
\newblock \bibinfo{journal}{\emph{International Journal of Information Management Data Insights}} \bibinfo{volume}{4}, \bibinfo{number}{1} (\bibinfo{year}{2024}), \bibinfo{pages}{100232}.
\newblock


\bibitem[Heng et~al\mbox{.}(2024)]%
        {heng2024comparing}
\bibfield{author}{\bibinfo{person}{Panha Heng}, \bibinfo{person}{Karn Yongsiriwit}, {and} \bibinfo{person}{Parkpoom Chaisiriprasert}.} \bibinfo{year}{2024}\natexlab{}.
\newblock \showarticletitle{Comparing the {Effectiveness} of {Generative} {AI} for {Learning} and {Developing} {Flutter} {Application}}. In \bibinfo{booktitle}{\emph{2024 8th {International} {Conference} on {Information} {Technology} ({InCIT})}}. \bibinfo{publisher}{IEEE}, \bibinfo{address}{Chonburi, Thailand}, \bibinfo{pages}{746--751}.
\newblock
\showISBNx{9798350366303}
\href{https://doi.org/10.1109/InCIT63192.2024.10810490}{doi:\nolinkurl{10.1109/InCIT63192.2024.10810490}}


\bibitem[Klotins et~al\mbox{.}(2023)]%
        {klotins2023continuous}
\bibfield{author}{\bibinfo{person}{Eriks Klotins}, \bibinfo{person}{Tony Gorschek}, {and} \bibinfo{person}{Magnus Wilson}.} \bibinfo{year}{2023}\natexlab{}.
\newblock \showarticletitle{Continuous software engineering: Introducing an industry readiness model}.
\newblock \bibinfo{journal}{\emph{IEEE Software}} \bibinfo{volume}{40}, \bibinfo{number}{4} (\bibinfo{year}{2023}), \bibinfo{pages}{77--87}.
\newblock


\bibitem[Le et~al\mbox{.}(2022)]%
        {le2022coderl}
\bibfield{author}{\bibinfo{person}{Hung Le}, \bibinfo{person}{Yue Wang}, \bibinfo{person}{Akhilesh~Deepak Gotmare}, \bibinfo{person}{Silvio Savarese}, {and} \bibinfo{person}{Steven Chu~Hong Hoi}.} \bibinfo{year}{2022}\natexlab{}.
\newblock \showarticletitle{Coderl: Mastering code generation through pretrained models and deep reinforcement learning}.
\newblock \bibinfo{journal}{\emph{Advances in Neural Information Processing Systems}}  \bibinfo{volume}{35} (\bibinfo{year}{2022}), \bibinfo{pages}{21314--21328}.
\newblock


\bibitem[Li et~al\mbox{.}(2023)]%
        {li2023large}
\bibfield{author}{\bibinfo{person}{Yinheng Li}, \bibinfo{person}{Shaofei Wang}, \bibinfo{person}{Han Ding}, {and} \bibinfo{person}{Hang Chen}.} \bibinfo{year}{2023}\natexlab{}.
\newblock \showarticletitle{Large language models in finance: A survey}. In \bibinfo{booktitle}{\emph{Proceedings of the Fourth ACM International Conference on AI in Finance}}. \bibinfo{pages}{374--382}.
\newblock


\bibitem[Nam et~al\mbox{.}(2024)]%
        {nam2024using}
\bibfield{author}{\bibinfo{person}{Daye Nam}, \bibinfo{person}{Andrew Macvean}, \bibinfo{person}{Vincent Hellendoorn}, \bibinfo{person}{Bogdan Vasilescu}, {and} \bibinfo{person}{Brad Myers}.} \bibinfo{year}{2024}\natexlab{}.
\newblock \showarticletitle{Using an llm to help with code understanding}. In \bibinfo{booktitle}{\emph{Proceedings of the IEEE/ACM 46th International Conference on Software Engineering}}. \bibinfo{pages}{1--13}.
\newblock


\bibitem[Omidvar~Tehrani et~al\mbox{.}(2024)]%
        {omidvar2024evaluating}
\bibfield{author}{\bibinfo{person}{Behrooz Omidvar~Tehrani}, \bibinfo{person}{Ishaani M}, {and} \bibinfo{person}{Anmol Anubhai}.} \bibinfo{year}{2024}\natexlab{}.
\newblock \showarticletitle{Evaluating {Human}-{AI} {Partnership} for {LLM}-based {Code} {Migration}}. In \bibinfo{booktitle}{\emph{Extended {Abstracts} of the {CHI} {Conference} on {Human} {Factors} in {Computing} {Systems}}}. \bibinfo{publisher}{ACM}, \bibinfo{address}{Honolulu HI USA}, \bibinfo{pages}{1--8}.
\newblock
\showISBNx{9798400703317}
\href{https://doi.org/10.1145/3613905.3650896}{doi:\nolinkurl{10.1145/3613905.3650896}}


\bibitem[Patton~Quinn(2002)]%
        {patton2002qualitative}
\bibfield{author}{\bibinfo{person}{Michael Patton~Quinn}.} \bibinfo{year}{2002}\natexlab{}.
\newblock \bibinfo{title}{Qualitative research \& evaluation methods}.
\newblock


\bibitem[Petersen and Wohlin(2009)]%
        {petersen2009context}
\bibfield{author}{\bibinfo{person}{Kai Petersen} {and} \bibinfo{person}{Claes Wohlin}.} \bibinfo{year}{2009}\natexlab{}.
\newblock \showarticletitle{Context in industrial software engineering research}. In \bibinfo{booktitle}{\emph{2009 3rd International Symposium on Empirical Software Engineering and Measurement}}. IEEE, \bibinfo{pages}{401--404}.
\newblock


\bibitem[Runeson and H{\"o}st(2009)]%
        {runeson2009guidelines}
\bibfield{author}{\bibinfo{person}{Per Runeson} {and} \bibinfo{person}{Martin H{\"o}st}.} \bibinfo{year}{2009}\natexlab{}.
\newblock \showarticletitle{Guidelines for conducting and reporting case study research in software engineering}.
\newblock \bibinfo{journal}{\emph{Empirical software engineering}} \bibinfo{volume}{14}, \bibinfo{number}{2} (\bibinfo{year}{2009}), \bibinfo{pages}{131}.
\newblock


\bibitem[Vaithilingam et~al\mbox{.}(2022)]%
        {vaithilingam2022expectation}
\bibfield{author}{\bibinfo{person}{Priyan Vaithilingam}, \bibinfo{person}{Tianyi Zhang}, {and} \bibinfo{person}{Elena~L Glassman}.} \bibinfo{year}{2022}\natexlab{}.
\newblock \showarticletitle{Expectation vs. experience: Evaluating the usability of code generation tools powered by large language models}. In \bibinfo{booktitle}{\emph{Chi conference on human factors in computing systems extended abstracts}}. \bibinfo{pages}{1--7}.
\newblock


\end{thebibliography}


\end{document}